\documentstyle[12pt,fleqn]{article}

\def\Slash#1{{\rm\ooalign{\hfil$/$\hfil\crcr \hbox{$#1$}}}}
\newcommand\Kmbf[1]{\mbox{\boldmath{$#1$}}}

\newtheorem{theorem}{Theorem}
\newcommand{\QED}{\hfill \rule[-1pt]{6pt}{6pt} \bigskip \par}

\begin{document}

\begin{center}
{\Large \bf
Approximation of the Schwinger--Dyson \par
and the Bethe--Salpeter Equations and Chiral Symmetry of QCD
} \par
\vskip 10mm
K. Naito\footnote{E-mail address: kenichi@th.phys.titech.ac.jp},
and M. Oka\\ 
{\it Department of Physics, Tokyo Institute of Technology,} \\
{\it Meguro, Tokyo 152-8551, Japan} 
\vskip 5mm
\end{center}
\vskip 5mm
\begin{abstract}
\baselineskip 1.5pc
 The Bethe--Salpeter equation for the pion in chiral
symmetric models is studied with a special care to consistency with 
low-energy relations.
We propose a reduction of the rainbow 
Schwinger--Dyson and the ladder Bethe--Salpeter equations
with a dressed gluon propagator.
We prove that the reduction preserves the Ward--Takahashi identity for the
axial-vector current and the PCAC relation.
\end{abstract}

\renewcommand{\thefootnote}{ \mbox{*}\arabic{footnote} }


\section{Introduction} \label{SEC:H100418:1}
 The Bethe--Salpeter(BS) equation is a popular approach to decribe
the pion 
as a relativistic bound state of a quark and an anti-quark. As the pion
is a pseudo Nambu--Goldstone(NG) boson, 
the results should satisfy 
the low-energy relations based on chiral symmetry. 
Therefore it is important to solve the Bethe--Salpeter equation
in keeping the low-energy relations.
The key idea is a consistency between the Schwinger--Dyson(SD)
equation for a quark and the BS equation.
If one solves the exact SD equation and 
the exact BS equation in QCD, the Ward--Takahashi identity
for the axial-vector current implies that 
the NG solution of the BS amplitude is given by
\begin{equation}
 \chi(q;P_B=0) = \frac{1}{f_\pi}\{i\gamma_5 
 \frac{\lambda^\alpha}{2},S_F(q) \}
 \label{AEQ:H100418:6}
\end{equation}
in the chiral limit, and that the exact relation
\begin{equation}
 M_\pi^2 f_\pi = -2m_0 \int_q {\rm tr}[\overline{\chi}(q;P_B) i\gamma_5\frac{\lambda^\alpha}{2}]
 \label{AEQ:H100418:8}
\end{equation}
holds for a finite quark mass $m_0$, where $S_F(q)$ is the quark propagator 
given as the solution of the SD
equation and $\chi(q;P_B)$ is the solution of the BS equation for the pion, 
\begin{eqnarray}
 S_F(q) & := & \int d^4 (x-y) e^{iq(x-y)} \langle 0 | T \psi(x)
 \overline{\psi}(y) | 0 \rangle, \label{AEQ:H100508:1} \\
 \chi(q;P_B) & := & e^{iP_BX} \int d^4(x-y) e^{iq(x-y)} \langle 0 | T 
 \psi(x) \overline{\psi}(y) | \Kmbf{P} \rangle,
 \label{AEQ:H100425:1}  \\
 \overline{\chi}(q;P_B) & := & e^{-iP_BX} \int d^4(y-x) e^{iq(y-x)} \langle
 \Kmbf{P} | T \psi(y) \overline{\psi}(x) | 0 \rangle.
 \label{AEQ:H100425:2}  
\end{eqnarray}
$|\Kmbf{P}\rangle$ denotes a pion state with normalization condition 
$\langle \Kmbf{P} | \Kmbf{P}'\rangle = (2\pi)^3 2 P_{B0} \delta^3(\Kmbf{P}-
\Kmbf{P}')$ and $P_{B\mu}:=(\sqrt{M_\pi^2+\Kmbf{P}^2},\Kmbf{P})$ denotes 
the on-shell momentum.
The pion decay constant $f_\pi$ is defined by 
\begin{equation}
 f_\pi := \lim_{P\to P_B} \frac{1}{P^2} \int_q {\rm tr}[ 
 \overline{\chi}(q;P) 
 i\gamma_5\frac{\lambda^\alpha}{2}\Slash{P}].
 \label{AEQ:H100418:7}
\end{equation}
From Eqs.$(\ref{AEQ:H100418:6})$ and $(\ref{AEQ:H100418:8})$,
the Gell-Mann, Oakes and Renner(GMOR) mass formula,
\begin{equation}
 M_\pi^2 f_\pi^2 \simeq -2m_0 \langle \overline{\psi}\psi\rangle_0 
 ,\quad
 \langle \overline{\psi}\psi \rangle_0 := 
 - \int_q {\rm tr}[S_F(q)_{m_0=0}]
 \label{AEQ:H100425:3}
\end{equation}
is derived for a small quark mass $m_0\simeq 0$. 
These relations are derived from the chiral symmetry  of
QCD and give an important starting point of the chiral
perturbation theory.

 In practice, the SD and BS equations can be solved only under some 
approximations. An effective model is often constructed and it is solved
with further truncations. Such approximations may not
always be consistent with chiral symmetry and the low-energy relations
are often violated.
It is known that 
if one uses the rainbow approximation for the SD equation
and the ladder approximation for the BS equation, 
the consistency between the SD and BS equations
is preserved and the low-energy relations
$(\ref{AEQ:H100418:6})$ and 
$(\ref{AEQ:H100418:8})$ still hold.
The consistency is also preserved after appropriate improvements
of the rainbow--ladder approximation
such as employing dressed gluon 
propagators\cite{MJ92,KM92,BQRTT97,MR97} 
in the interaction kernel.\cite{Mun95,NYNOT99}

For phenomenological studies, however, 
the ladder BS equation is rather involved numerically  
and therefore further approximations are often used.
In this case the consistency between the SD and 
BS equations may be lost and there is no guarantee for
the low-energy relations.
In this paper, we propose an approximation taking only dominant
terms in the SD and BS equations.
We prove that the approximation preserves the low-energy relations 
$(\ref{AEQ:H100418:6})$ and $(\ref{AEQ:H100418:8})$, while it may change 
values of $M_\pi,\,f_\pi$ 
and $\langle\overline{\psi}\psi\rangle$.


\section{Approximation of the SD and BS Equations} \label{SEC:H100418:2}
 In this paper we concentrate on the rainbow SD equation for the quark 
propagator,
\begin{equation}
 S_F^{-1}(q) = S_0^{-1}(q) + i 
 C_F \int_k g^2((q-k)^2) iD^{\mu\nu}(q-k)\gamma_\mu 
 S_F(k)\gamma_\nu \label{AEQ:H100418:1}
\end{equation}
and the ladder BS equation for the pion
\begin{eqnarray}
  S_F^{-1}(q_{+B}) \chi(q;P_B) S_F^{-1}(q_{-B}) 
 =  -i C_F \int_k g^2((q-k)^2) 
 iD^{\mu\nu}(q-k)\gamma_\mu \chi(k;P_B) \gamma_\nu ,
 \label{AEQ:H100418:2}
\end{eqnarray}
\begin{equation}
 q_{\pm B} := q \pm \frac{P_B}{2},
 \label{AEQ:H100419:5}
\end{equation}
where $S_0(q)$ is the bare quark propagator defined by
\begin{equation}
 S_0(q) := \frac{i}{\Slash{q}-m_0}. \label{AEQ:H100418:3}
\end{equation}
The normalization condition of the BS amplitude is given by
\begin{equation}
 \lim_{P\to P_B} 
 \frac{-P^\mu}{2P^2} i\int_q
\overline{\chi}_{m_2n_2}(q;P_B) 
\frac{\partial}{\partial P^\mu}\Big(
S^{-1}_{Fn_2n_1}(q_{+B}) S^{-1}_{Fm_1m_2}(q_{-B}) \Big) 
 \chi_{n_1m_1}(q;P_B) = 1.
\label{AEQ:H100419:1}
\end{equation}
The indices $m,n,\cdots$
are combined indices $m:=(a,i,f),\,n:=(b,j,g),\cdots$ with Dirac indices
$a,b,\cdots$ and color indices $i,j,\cdots $ and flavor indices $f,g,\cdots$.
In Eqs.$(\ref{AEQ:H100418:1})$ and $(\ref{AEQ:H100418:2})$,
$g^2((q-k)^2)$ is a running coupling constant and $iD^{\mu\nu}(q-k)$
denotes a dressed gluon propagator, which is modified due to 
perturbative and/or non-perturbative corrections.
For simplicity, we do not consider divergence from quark loops nor 
renormalization. 

 The above set of the SD and BS equations has been studied
by many authors in Refs.\cite{MJ92,KM92,BQRTT97,MR97,RW94}
with successes in phenomenology.
However, as far as we know, the full calculation of
the ladder BS equation for finite $m_0$ is reported
only in Ref.\cite{MR97}.
Because the BS equation requires a large computation, the studies
often make approximations, such as neglecting 
some terms and/or truncating the Chebyshev polynomial
expansion.
But these approximations of the BS equation 
are often inconsistent with the SD equation, and therefore
the low-energy relations are violated,
although the numerical result in Ref.\cite{MR97} shows that 
the violation may not be so large.

 In general, the solution of the SD equation is written in terms of
two scalar functions $A(q^2)$ and $B(q^2)$ as
\begin{equation}
 S_F(q) = \frac{i}{A(q^2)\Slash{q}-B(q^2)}
 \label{AEQ:H100418:4}
\end{equation}
and the solution of the BS equation is written in terms of four scalar
functions $\phi_S(q;P_B),\,\phi_P(q;P_B),\,$ $\phi_Q(q;P_B)$ and
$\phi_T(q;P_B)$ as
\begin{eqnarray}
 \lefteqn{ \chi_{nm}(q;P_B) = \delta_{ji} \frac{\lambda^\alpha_{gf}}{2}\bigg[
 \bigg(\phi_S(q;P_B) + \phi_P(q;P_B) \Slash{q} + \phi_Q(q;P_B)\Slash{P}_{\!B}
  }\nonumber \\
 & & {} + \frac{1}{2} \phi_T(q;P_B)(\Slash{P}_{\!B}\Slash{q}-\Slash{q}\Slash{P}_{\!B} )\bigg)\gamma_5\bigg]_{ba}.
 \label{AEQ:H100418:5}
\end{eqnarray}

 The following theorem addresses the consistency of the rainbow SD
and the ladder BS equations with chiral symmetry:
\begin{theorem} \label{TH:H100420:1}
  If one solves the rainbow SD equation $(\ref{AEQ:H100418:1})$ and the 
ladder BS equation $(\ref{AEQ:H100418:2})$ with the 
normalization condition $(\ref{AEQ:H100419:1})$, the low-energy relations
$(\ref{AEQ:H100418:6})$ and $(\ref{AEQ:H100418:8})$ hold.
\end{theorem}
This is known as mutual consistency between the SD and BS 
equations.\cite{Mun95,NYNOT99}. If one uses a further approximation 
inconsistent between the SD and BS equations,
the low-energy relations may be violated.

 We here propose an approximation which is consistent with chiral
symmetry: 
\begin{theorem} \label{TH:H100420:2}
 If one makes the approximation 
$A(q^2)\equiv 1$ in the SD equation $(\ref{AEQ:H100418:1})$ and 
neglects the $\phi_P(q;P_B),\phi_Q(q;P_B),\phi_T(q;P_B)$ terms in RHS of 
the BS equation
$(\ref{AEQ:H100418:2})$, the low-energy relations $(\ref{AEQ:H100418:6})$ and
$(\ref{AEQ:H100418:8})$ still hold.
\end{theorem}
 This approximation of the BS equation does not imply to neglect 
$\phi_P(q;P_B),\,\phi_Q(q;P_B),\,\phi_T(q;P_B)$ in LHS of 
Eq.$(\ref{AEQ:H100418:2})$ or in Eq.$(\ref{AEQ:H100419:1})$.
Instead they are given by 
\begin{eqnarray}
 \phi_P(q;P_B) & = & \frac{ B(q_{-B})-B(q_{+B}) }{-q^2 + P_B^2/4 + B(q_{-B}) B(q_{+B})} \phi_S(q;P_B), \label{AEQ:H100419:2} \\
 \phi_Q(q;P_B) & = & \frac{ B(q_{-B})+B(q_{+B}) }{2(-q^2 + P_B^2/4 + B(q_{-B}) B(q_{+B}))} \phi_S(q;P_B), \label{AEQ:H100419:3} \\
 \phi_T(q;P_B) & = & \frac{ -1}{-q^2+P_B^2/4 + B(q_{-B}) B(q_{+B}) } \phi_S(q;P_B). \label{AEQ:H100419:4}
\end{eqnarray}

 Before the proof of this theorem, 
we review how to derive the SD and BS equations.
We use the Cornwall--Jackiw--Tomboulis(CJT) effective action formulation.\cite{CJT74,Mir93}
In the rainbow--ladder approximation, the CJT action is given by
\begin{equation}
 \Gamma[S_F] = i{\rm Tr}{\rm Ln}[S_F] - i{\rm Tr}[S^{-1}_0 S_F] 
 + \Gamma_{\rm loop}[S_F],
 \label{AEQ:H100419:6}
\end{equation}
\begin{equation}
 \Gamma_{\rm loop}[S_F] = \frac{1}{2} \int d^4x {\cal K}^{m_1m_2,n_1n_2}
 (i\partial_{x_1},i\partial_{x_2};i\partial_{y_1},i\partial_{y_2})
 S_{Fm_2n_1}(x_2,y_1)S_{Fn_2m_1}(y_2,x_1)] |_*
 \label{AEQ:H100419:7}
\end{equation}
where the symbol $*$ means to take $x_1,x_2,y_1,y_2 \to x$ after all
the derivatives are operated. ${\cal K}$ denotes the
 interaction kernel defined by
\begin{eqnarray}
\lefteqn{ {\cal K}^{m_1m_2,n_1n_2}(p_1,p_2;q_1,q_2) := g^2( 
(\frac{p_1+p_2-q_1-q_2}{2})^2 ) } \nonumber \\
& & \times 
 iD^{\mu\nu}(\frac{p_1+p_2}{2}-\frac{q_1+q_2}{2})(\gamma_\mu T^a)^{m_1m_2}
 (\gamma_\nu T^a)^{n_1n_2}. \label{AEQ:H100419:8}
\end{eqnarray}
 The SD equation $(\ref{AEQ:H100418:1})$ is the stability condition of the CJT action
\begin{equation}
 \frac{\delta \Gamma[S_F]}{\delta S_{Fmn}(x,y)} = 0,
 \label{AEQ:H100420:2}
\end{equation}
and the (homogeneous) BS equation $(\ref{AEQ:H100418:2})$ 
and its normalization condition
 $(\ref{AEQ:H100419:1})$ are derived from the inhomogeneous BS equation
\begin{equation}
 \frac{1}{i} \frac{\delta ^2 \Gamma[S_F]}{\delta S_{Fmn}(x,y) \delta
  S_{Fm'n'}(y',x')} G_{C;n'm'm''n''}^{(2)}(y'x';x''y'') =
   \delta_{m''m}\delta_{nn''}\delta(x''-x)\delta(y-y'')
 \label{AEQ:H100419:9}
\end{equation}
where the repeated indices are summed or intergrated and 
 $G_{C;n'm'm''n''}^{(2)}(y'x';x''y'')$ is the two-body 
connected Green function defined by
\begin{eqnarray}
 G_{C;nmm'n'}^{(2)}(yx;x'y') & := & \langle 0 | T \psi_n(y)
 \overline{\psi}_m(x)
 \psi_{m'}(x') \overline{\psi}_{n'}(y') | 0 \rangle  \nonumber \\
& & {} \quad - \langle 0 | T\psi(y)_n\overline{\psi}_m(x) |
 0 \rangle \langle 0 | T\psi_{m'}(x')\overline{\psi}_{n'}(y')
  | 0 \rangle. \label{AEQ:H100419:10}
\end{eqnarray}
Especially, the form of the (homogeneous) BS equation is
given by
\begin{equation}
\frac{\delta^2 \Gamma[S_F]}{\delta S_{Fmn}(x,y)\delta S_{Fn'm'}(y',x')}
\chi_{n'm'}(y',x';P_B) = 0.
\label{AEQ:H100420:3}
\end{equation}

The SD and BS equations inherit the chiral property of 
the CJT action Eq.$(\ref{AEQ:H100419:6})$.\cite{Mun95}
It is easy to show that if we apply an approximation that does not violate
(global) chiral symmetry directly to the CJT action, then
the SD and BS equations resulting from the approximated CJT action
preserve chiral symmetry.
It is, however, shown that non-trivial momentum dependences in the 
approximation may result in changing the axial-vector Noether current
and thus in modifying the low-energy relations
$(\ref{AEQ:H100418:6})$ and 
$(\ref{AEQ:H100418:8})$.\cite{NYNOT99}
In order to keep the low-energy relations in the original 
forms $(\ref{AEQ:H100418:6})$ and 
$(\ref{AEQ:H100418:8})$, the interaction part of the CJT action 
$\Gamma_{\rm loop}[S_F]$ must be invariant
under the local axial infinitesimal transformation
\begin{equation}
 S_F(x,y) \to S_F'(x,y) := (1+i\gamma_5\frac{\lambda^\alpha}{2}\theta^\alpha(x))S_F(x,y) (1+i\gamma_5\frac{\lambda^\alpha}{2}\theta^\alpha(y)).
 \label{AEQ:H100419:11}
\end{equation}
Indeed, it can be shown that the rainbow--ladder 
approximation $(\ref{AEQ:H100419:7})$ and $(\ref{AEQ:H100419:8})$ 
leaves $\Gamma_{\rm loop}[S_F]$ invariant. 

 The invariance 
of $\Gamma_{\rm loop}[S_F]$ under the transformation 
Eq.$(\ref{AEQ:H100419:11})$ leads us to an equation
\begin{eqnarray}
\lefteqn{ G^{(2)}_{C;m''n''nm}(x''y'';yx)\frac{\delta \Big( \Delta_5
\Gamma[S_F] \Big)}{\delta S_{Fnm}(y,x)}   } \nonumber \\
& \equiv &
 G^{(2)}_{C;m''n''nm}(x''y'';yx) 
 \frac{\delta }{\delta S_{Fnm}(y,x)} \left\{
\frac{\delta \Gamma[S_F]}{\delta S_{Fn'm'}
 (y',x')}\{i\gamma_5 \frac{\lambda^\alpha}{2}\theta^\alpha,S_F\}_{n'm'
 }(y',x') \right\} \nonumber \\
& \equiv & G^{(2)}_{C;m''n''nm}(x''y'';yx) \frac{\delta^2\Gamma[S_F]}{
\delta S_{Fnm}(y,x)\delta S_{Fm'n'}(x',y')}\{i\gamma_5\frac{\lambda^\alpha
}{2}\theta^\alpha,S_F\}_{m'n'}(x',y') \nonumber \\
& & {}+ \frac{\delta \Gamma[S_F]}{\delta S_{Fm'n'}(x',y')}\bigg\{ G^{(2)
}_{C;m''n''m'l}(x''y'';x'y')(i\gamma_5\frac{\lambda^\alpha}{2}
\theta^\alpha(y'))_{ln'} \nonumber \\
& & {} \quad +(i\gamma_5\frac{\lambda^\alpha}{2}\theta^\alpha(x')
)_{m'l}G^{(2)}_{C;m''n''ln'}(x''y'';x'y')\bigg\} \nonumber
 \\ 
& = & i\{i\gamma_5\frac{\lambda^\alpha}{2}\theta^\alpha,S_F\}_{m''n''}(x'',y''),
 \label{AEQ:H100430:1}
\end{eqnarray}
where the SD equation $(\ref{AEQ:H100420:2})$
and the inhomogeneous BS equation $(\ref{AEQ:H100419:9})$ are used
for the last equality.
In the momentum space Eq.$(\ref{AEQ:H100430:1})$ becomes
\begin{eqnarray}
\lefteqn{ \int_q G^{(2)}_{C;m''n''nm}(p,q;P)\Big( i\gamma_5 
\frac{\lambda^\alpha}{2}(2m_0+\Slash{P})\Big)_{mn} } \nonumber \\
& = & i\Big( i\gamma_5\frac{\lambda^\alpha}{2}S_F(p-\frac{P}{2})\Big)_{m''n''}
 + i\Big( S_F(p+\frac{P}{2})i\gamma_5\frac{\lambda^\alpha}{2}\Big)_{m''n''}.
 \label{AEQ:H100430:2}
\end{eqnarray}
 Here the Fourier transformation of $G^{(2)}_{C;m'n'nm}(x'y';yx)$
is defined by
\begin{equation}
 G^{(2)}_{C;m'n'nm}(x'y';yx) = \int_{pqP}e^{-i\{p(x'-y')+q(y-x)+P(X'-X)\}}
 G^{(2)}_{C;m'n'nm}(p,q;P).
 \label{AEQ:H100430:3}
\end{equation}
Note that LHS of Eq.$(\ref{AEQ:H100430:2})$ contains contributions
only from the second term of the CJT action $(\ref{AEQ:H100419:6})$,
because $\Gamma_{\rm loop}[S_F]$ is invariant under the transformation 
Eq.$(\ref{AEQ:H100419:11})$.
Then the low-energy relations $(\ref{AEQ:H100418:6})$ and 
$(\ref{AEQ:H100418:8})$ are derived from Eq.$(\ref{AEQ:H100430:2})$ 
and the spectral
representation of $G^{(2)}_{C;m'n'nm}(p,q;P)$
\begin{equation}
 G^{(2)}_{C;nmm'n'}(yx;x'y')  =  \int_P  e^{-i(P-P_B)X+i(P-P_B)X'}  i\frac{\chi_{nm}(y,x;P_B)\overline{\chi}_{m'n'}(x',y';P_B)}
{P^2-M_\pi^2+i\epsilon} + \cdots \label{AEQ:H100430:4}
\end{equation}
with
\begin{equation}
 X := \frac{ x + y}{2} ,\quad X' := \frac{ x' + y'}{2},\quad
 P=(P_0,\Kmbf{P}),\quad P_B=(\sqrt{M_\pi^2+\Kmbf{P}^2},\Kmbf{P}).
 \label{AEQ:H100430:6}
\end{equation}
(See section {\bf 3.1} in Ref.\cite{NYNOT99}.)

 Now we prove the theorem \ref{TH:H100420:2}.
 Consider a projection operator,
\begin{equation}
 {\cal P}_{ba,a'b'} := 
 \frac{1}{4}\Big[ (1)_{ba} (1)_{a'b'} 
 + (\gamma_5)_{ba} (\gamma_5)_{a'b'} \Big],
  \label{AEQ:H100419:12}
\end{equation}
which satisfies
\begin{equation}
 {\cal P}_{ba,a'b'}(  S +  \gamma_5 P +  \gamma_\mu V  
 + \gamma_\mu\gamma_5 A+ \gamma_\mu\gamma_\nu T )_{b'a'}
   = (S + \gamma_5 P)_{ba}.
  \label{AEQ:H100419:13}
\end{equation}
Using this operator, we modify the interaction kernel as
\begin{eqnarray}
\lefteqn{
 {\cal K}^{m_1m_2,n_1n_2}(p_1,p_2;q_1,q_2) \to 
 {\cal K}_{\cal P}^{m_1m_2,n_1n_2}(p_1,p_2;q_1,q_2) } \nonumber \\
 & := & \delta_{i_1i_1'}\delta_{i_2'i_2}\delta_{j_1j_1'}\delta_{j_2'j_2}
 \delta_{f_1f_1'}\delta_{f_2f_2'}\delta_{g_1g_1'}\delta_{g_2'g_2} 
 {\cal P}_{b_2a_1,a_1'b_2'} {\cal P}_{a_2b_1,b_1'a_2'}
 {\cal K}^{m_1'm_2',n_1'n_2'} \label{AEQ:H100419:14}
\end{eqnarray}
in Eq.$(\ref{AEQ:H100419:7})$.
This modified kernel ${\cal K}_{\cal P}$ has the properties 
\begin{eqnarray}
 {\cal K}_{\cal P}^{mm',nl}(p,p';q,q') (i\frac{\lambda^\alpha}{2})_{ln'} & = &
 (i\frac{\lambda^\alpha}{2})_{nl} {\cal K}_{\cal P}^{mm',ln'}(p,p';q,q'), 
 \label{AEQ:H100420:5} \\
 {\cal K}_{\cal P}^{mm',nl}(p,p';q,q') (i\gamma_5\frac{\lambda^\alpha}{2})_{ln'} 
 & = &  -(i\gamma_5\frac{\lambda^\alpha}{2})_{nl} 
 {\cal K}_{\cal P}^{mm',ln'}(p,p';q,q'), 
 \label{AEQ:H100420:6}
\end{eqnarray}
\begin{equation}
 {\cal K}_{\cal P}^{mm',nn'}(p,p';q,q') = 
 {\cal K}_{\cal P}^{mm',nn'}(p+p';q+q').
 \label{AEQ:H100308:8}
\end{equation}
These properties imply that the interaction term, $\Gamma_{\rm loop}[S_F]$,
is invariant under 
not only the global but also the local chiral
transformation Eq.$(\ref{AEQ:H100419:11})$.
 Therefore this modification preserves
the low-energy relations $(\ref{AEQ:H100418:6})$ and 
$(\ref{AEQ:H100418:8})$.\cite{NYNOT99}

 To complete the proof, we show that 
this modification of the interaction kernel 
corresponds to the approximation stated 
in the theorem \ref{TH:H100420:2}.
Under the approximation $(\ref{AEQ:H100419:14})$,
the SD equation $(\ref{AEQ:H100418:1})$ is modified to
\begin{equation}
 S_F^{-1}(q) = S_0^{-1}(q) + 
 i\frac{C_F}{4} \int_k g^2((q-k)^2) iD^{\mu\nu}(q-k)
 {\rm tr}^{\rm (D)}[S_F(k)] \gamma_\mu 
 \gamma_\nu, \label{AEQ:H100420:1}
\end{equation}
where ${\rm tr}^{\rm (D)}$ denotes the trace in the
Dirac space.
As the second term of RHS in Eq.$(\ref{AEQ:H100420:1})$ contains only 
the scalar component, $A(q^2)$ is 
identically $1$. Thus Eq.$(\ref{AEQ:H100420:1})$ is equivalent to
Eq.$(\ref{AEQ:H100418:1})$ with the condition 
$A(q^2)\equiv 1$.
Similarly, the BS equation $(\ref{AEQ:H100418:2})$ 
is modified to
\begin{eqnarray}
\lefteqn{  S_F^{-1}(q_{+B}) \chi(q;P_B) S_F^{-1}(q_{-B})  }
\nonumber \\
& = & -i\frac{C_F}{4}\int_k g^2((q-k)^2) iD^{\mu\nu}(q-k)
  {\rm tr}^{\rm (D)}[\chi(k;P_B)\gamma_5] 
  \gamma_\mu \gamma_5 \gamma_\nu
 \label{AEQ:H100420:4}
\end{eqnarray}
in which RHS contains only the pseudo-scalar component.
This corresponds to neglecting $\phi_P(k;P),\phi_Q(k;P)$ and
$\phi_T(k;P)$ in RHS of Eq.$(\ref{AEQ:H100418:2})$.
The proof is ended.\QED
 Theorem \ref{TH:H100420:2} gives a consistent way of reducing
the number of degrees of the SD and BS equations and therefore
enables us to save computer time. Because the deviation
of $A(q^2)$ from unity is usually small and also $\phi_S(q;P_B)$
is the main term of the pion BS amplitude, this approximation
will be useful in phenomenology. As this theorem guarantees that
the low-energy relations $(\ref{AEQ:H100418:6})$ and 
$(\ref{AEQ:H100418:8})$ are satisfied, properties of chiral symmetry and
its spontaneous breaking may be studied in this approximation.
It is, however, noted that this approximation is regarded as modification
of the model interaction as seen in Eqs.$(\ref{AEQ:H100419:13})$
and $(\ref{AEQ:H100419:14})$. 
As the projection $(\ref{AEQ:H100419:13})$ truncates 
the interaction kernel to the scalar and pseudo-scalar components,
the resulting model looks very similar to the (non-localized)
Nambu--Jona-Lasinio model. This indicates that the approximation
may be valid for the pion, pseudo-scalar meson, but it may give 
large modification to the vector mesons, for instance.
Thus we should consider that the approximation gives a different
model, while it gives good approximation to the pseudo-scalar
channel.

 Further we notice the following theorem, which may sometimes be useful:
\begin{theorem} \label{TH:H100423:1}
 If one uses the approximation that
neglects only the $\phi_T(q;P_B)$ term in RHS of 
the BS equation
$(\ref{AEQ:H100418:2})$, the low-energy relations $(\ref{AEQ:H100418:6})$ and
$(\ref{AEQ:H100418:8})$ are satisfied.
\end{theorem}
This can be proved similarly by using the projection operator
\begin{equation}
 {\cal P'}_{ba,a'b'} := \frac{1}{4}
 \Big[ (1)_{ba} (1)_{a'b'}+(\gamma_5)_{ba} (\gamma_5)_{a'b'}
 + (\gamma_\mu)_{ba}(\gamma^\mu)_{a'b'}
 + (\gamma_5\gamma_\mu)_{ba}(\gamma^\mu \gamma_5)_{a'b'}\Big].
 \label{AEQ:H100425:4}
\end{equation}

\section{Summary and Conclusions} \label{SEC:H100423:1}
 In this paper, we have proposed a new approximation scheme of the SD and
BS equations. In general, approximations violate the 
low-energy properties originated from chiral symmetry.
 To avoid this, one must include many independent degrees in equations,
although most of them are not interesting from
the physical view-point and
their contributions are not so large. We have proved
that the low-energy relations are preserved when one takes the
scalar term of the SD equation and the psuedo-scalar 
term of the BS equation. This approximation is consistent with the
picture that 
the dynamical mass generation of the
quark is described mainly by the scalar part in the SD equation,
while the bound state structure of the pion BS (amputated) amplitude  
is composed mainly of pseudo-scalar term accordingly. 
In Ref.\cite{FR96,CG96,LK96,CG97}, the GMOR relation and its improvement
were studied, where the BS equation is approximated by taking only the
pseudo-scalar term although 
the SD solution contains both the vector and the scalar parts.
This corresponds to using the projection operator
\begin{equation}
 {\cal P}_{ba,a'b'} := 
 \frac{1}{4}\Big[ (\gamma_5)_{ba} (\gamma_5)_{a'b'} \Big]
  \label{AEQ:H100507:1}
\end{equation}
instead of Eq.$(\ref{AEQ:H100419:12})$ and violates not only the
local but also the global chiral invariance of $\Gamma_{\rm loop}[S_F]$.
As a result the low-energy relations $(\ref{AEQ:H100418:6})$ and 
$(\ref{AEQ:H100418:8})$ may not be satisfied.

If they also approximate the SD equation by
imposing $A(q^2)\equiv 1$, then chiral symmetry 
may be recovered. We have also proposed another approximation in 
which the tensor term of the BS equation is neglected.
It is worth mentioning that the 
tables I and II in Ref.\cite{MR97} confirms 
our theorem \ref{TH:H100423:1} numerically.

 So far we have concentrated on the effective interaction of type 
$(\ref{AEQ:H100419:8})$, which is based on the exchange of a dressed
gluon. But it is not so difficult to apply
the same approach to other types of the effective interaction.  
In Ref.\cite{NYNOT99}, we studied an effective model in which 
the axial-vector current is modified.
The application to such a model is obvious, for example.
 We hope that our approximation scheme helps analyses of
the chiral effective theories of QCD.


\section*{Acknowledgement}
 The authors would like to thank Drs. 
M. Takizawa, K. Yoshida, Y. Nemoto, A. G. Williams, A. Bender 
and K. Kusaka for valuable discussions. 
 This work is supported in part by the Grant-in-Aid
for scientific research (A)(1) 08304024 and (C)(2) 08640356
of the Ministry of Education, Science and Culture
of Japan.

\end{document}